\UseRawInputEncoding

\documentclass[sn-mathphys,Numbered]{sn-jnl}

\usepackage{graphicx}%
\usepackage{multirow}%
\usepackage{amsmath,amssymb,amsfonts}%
\usepackage{amsthm}%
\usepackage{mathrsfs}%
\usepackage[title]{appendix}%
\usepackage{xcolor}%
\usepackage{textcomp}%
\usepackage{manyfoot}%
\usepackage{booktabs}%
\usepackage{algorithm}%
\usepackage{algorithmicx}%
\usepackage{algpseudocode}%
\usepackage{listings}%
\usepackage{pdflscape}

\usepackage{array}

\theoremstyle{thmstyleone}%

\theoremstyle{thmstyletwo}%

\theoremstyle{thmstylethree}%

\begin{document}

\title[Article Title]{Benchmarking Large Language Models for Log Analysis, Security, and Interpretation}

\author*[1]{\fnm{Egil} \sur{Karlsen}}\email{egil.karlsen@dal.ca}

\author[2]{\fnm{Xiao} \sur{Luo}}\email{xiao.luo@okstate.edu}

\author[1]{\fnm{Nur} \sur{Zincir-Heywood}}\email{zincir@dal.ca}

\author[1]{\fnm{Malcolm} \sur{Heywood}}\email{mheywood@cs.dal.ca}

\affil[1]{\orgdiv{Faculty Computer Science}, \orgname{Dalhousie University}, \orgaddress{\street{University Ave.}, \city{Halifax}, \postcode{B3H 1W5}, \state{Nova Scotia}, \country{Canada}}}

\affil[2]{\orgdiv{Department of Management Science and Information Systems}, \orgname{Oklahoma State University}, \orgaddress{\street{370 Business Building}, \city{Stillwater}, \postcode{74078}, \state{Oklahoma}, \country{USA}}}

\abstract{Large Language Models (LLM) continue to demonstrate their utility in a variety of emergent capabilities in different fields. An area that could benefit from effective language understanding in cybersecurity is the analysis of log files. This work explores LLMs with different architectures (BERT, RoBERTa, DistilRoBERTa, GPT-2, and GPT-Neo) that are benchmarked for their capacity to better analyze application and system log files for secruity. Specifically, 60 fine-tuned languagemodels for log analysis are deployed and bechmarked. The resulting models demonstrate that they can be used to perform log analysis effectively with fine-tuning being particularly important for appropriate domain adaptation to specific log types. The best-performing fine-tuned sequence classification model (DistilRoBERTa) outperforms the current state-of-the-art; with an average F1-Score of 0.998 across six datasets from both web application and system log sources. To achieve this, we propose and implement a new experimentation pipeline (LLM4Sec) which leverages LLMs for log analysis experimentation, evaluation, and analysis.
}


\maketitle

\section{Introduction}\label{sec1}

In today's digital landscape, networks, especially within enterprise infrastructure, are extensive and active 24/7, often catering to hundreds of thousands of users concurrently. This continuous activity implies that log files are generated at the same rate by applications, services, systems, and networks; where log files are integral to cybersecurity monitoring, forensics, and alert mechanisms. With high user traffic, these logs can comprise hundreds of millions of records from diverse architectural layers. Traditionally, these voluminous and varied logs are aggregated via centralized systems like SIEMs, facilitating log analysis and anomaly detection.

Conventional log analysis relies on static regular expression parsers (e.g. DRAIN \cite{Drain}) to extract salient features from the aggregated logs. However, this methodology poses significant constraints, primarily because parsers must be preconfigured for specific log formats, and given precise instructions on the useful features that should be extracted. This is where Natural Language Processing (NLP) techniques offer substantial advantages. NLP enables dynamic feature extraction without the prerequisite of prior knowledge about data structures, enhancing scalability and, crucially, versatility \cite{OG, Italians}.

Existing research underscores the potential of NLP in application, network, service, and system log analysis. Studies indicate that semantic information from word2vec models, coupled with syntactic insights from TF-IDF, can yield meaningful vector space representations \cite{Italians}. Moreover, Karlsen et al. have demonstrated the advantage of Large Language Models (LLMs)-based semantic feature extraction over traditional syntactic methods in analyzing application logs for security analysis \cite{CSNet}. Specifically, their research highlights the efficacy of employing DistilRoBERTa, a prominent LLM, for generating sentence embeddings from raw log data, by evaluating its ability to disambiguate abnormal events using unsupervised machine learning methods.

Despite these advancements, the exploration of NLP-based log analysis, especially through LLMs, has room to improve. This study aims to bridge the gap by delving into an LLM-based approach for feature extraction in log analysis. It benchmarks various LLMs across application, system, and network-level log datasets, evaluating the approach's versatility for understaing anomolous behaviour. Furthermore, this study investigates the benefits of domain adaptation via the fine-tuning of LLMs.

Building on the earlier work of Karlsen et al., this research employs five distinct LLM architectures, evaluated using six labelled log datasets for security. It also evaluates the advantages of domain adaptation through the individual fine-tuning of LLMs using sequence classification, a supervised classification method. The contributions of this research are as follows:

\begin{itemize}
    \item \textbf{Evaluate Five LLMs}: Assess the efficacy of five distinct Large Language Models in identifying abnormal events and analyzing application and system logs, thereby establishing a performance benchmark in the field.
    \item \textbf{Demonstrate Versatility}: Confirm the adaptability and broad applicability of LLM-based approaches through tests across six log files.
    \item \textbf{Optimize with Fine-Tuning}: Explore the impact of advanced fine-tuning on LLMs using sequence classification, to enhance log analysis precision.
    \item \textbf{Boost Interpretability}: Employ visualizations and case studies to deepen the understanding and interpretation of the LLM-based approach in log analysis.
    \item \textbf{Provide Experimental pipeline}: Develop "LLM4Sec," a pipeline for experimenting with LLM-based log analysis for security.
\end{itemize}

The remaining sections of this paper are organized as follows: Section 2 reviews the literature in the field. Section 3 introduces the proposed approach and discusses the datasets, feature extraction, domain adaptation, and supervised learning models used in this research. Section 4 presents the evaluations and the results obtained. Finally, conclusions are drawn and future work is discussed in Section 5.

\section{Literature Review}\label{LR}
In this section, we give an overview of previous work on log analysis via NLP and LLM-based approaches.

\subsection{Traditional NLP Approaches}

Nguyen et al. \cite{supe} propose a supervised ensemble learning system that combines outputs from various supervised machine learning models to optimize abnormal security event classification. They test this pipeline on the ECML/PKDD 2007 and HTTP CSIC 2010 datasets using hand-picked features and log line syntactic characteristics. They emphasize the importance of adapting intrusion detection systems (IDSs) to heterogeneous and adversarial network environments. The proposed IDS outperforms baseline supervised machine learning methods by 10\% in accuracy, achieved through an online learning pipeline. The IDS incorporates supervised machine learning algorithms, including decision stump, Naive Bayes, RBF Network, and Bayesian Network. A significant contribution is a novel pipeline for fusing IDS outputs to enhance accuracy and reliability. The IDS achieves an accuracy of 0.9098 in the CSIC dataset and 0.9256 in the PKDD/ECML dataset, albeit with higher computational costs for model output fusion.

Vartouni et al. \cite{moradi2019leveraging} explore novel feature extraction techniques for intrusion detection using deep belief networks and parallel feature fusion methods. They apply synthetic feature extraction and feature fusion to bi-gram representations of log files, alongside dimensionality reduction algorithms like FICA, PCA, and KPCA. Their evaluation accesses the impacts of performing dimensionality reduction, synthetic feature fusion, and baseline n-gram representations for training unsupervised anomaly detection algorithms (Isolation forest, Elliptic envelope, and one-class SVM). Benchmarking is conducted using the ECML/PKDD 2007 and HTTP CSIC datasets. Results suggest that deep learning neural networks, especially fusion models, provide significant advantages in feature extraction for anomaly detection. Notably, the feature fusion method achieves an accuracy of 0.8535 for Isolation forest in the CSIC dataset and 0.8493 for Elliptic envelope in the PKDD dataset after extensive exploration and parameter tuning.

Bhatnagar et al. \cite{supe2} benchmark the performance of four supervised machine learning algorithms for intrusion detection. Their solution extracts features from raw log files using classical neural networks and deep belief networks. The top five highest entropy features are selected from the ECML/PKDD 2007 and HTTP CSIC 2010 datasets and a bag-of-words vector space complements these features. This hybrid feature set trains four supervised classifiers (MLP classifier, Multinomial classifier, decision tree, and random forest), all achieving an average accuracy of 0.996 in identifying attacks across both datasets.

Farzad et al. \cite{farzad} propose a supervised approach to anomaly detection which explores the use of word frequency in security log characteristics in combination with autoencoders and GRU, LSTM, and Bi-LSTM for anomaly detection. There are two auto-encoders in their proposed method, individually trained on negative (anomalous) and positively (normal) labelled log word frequency data with the aim of extracting meaningful features toward either class. The success of the autoencoders in extracting meaningful representations of the preprocessed log data is evaluated using deep learning methods GRU, LSTM, and Bi-LSTM. They name the different classification methods Auto-GRU, Auto-LSTM, and AutoBLSTM respectively, and evaluate on publicly availble system log files. 

Sivri et al. \cite{sivri} experimented with a variety of different machine learning approaches applied to character features from the ECML/PKDD 2007 and HTTP CSIC 2010 datasets. They pre-process the datasets by extracting handpicked log features and converting these features to lowercase UTF-8 encodings. They identify that the LSTM deep learning approach has the best classification performance in intrusion detection applied to the character-level representations. This work requires careful selection and extraction of features which makes the approach difficult to apply to other log formats.

Adhikari et al. \cite{adhikari} experiment with carefully selected log features from the HTTP CSIC 2010 dataset which range from statistical log line characteristics like character count to request type. They explore these feature representations and evaluate their ability to aid in intrusion detection when combined with several machine learning techniques, where XGBoost achieves the best results. This methodology also relies on carefully selected features and as such also suffers from limited adaptability for different log formats.

Copstein et al. \cite{copstein} investigate syntactic features based on the TF-IDF technique for intrusion detection in application log files. They use TF-IDF to rank terms based on uniqueness in the entire dataset, suggesting this feature set as a means to extract the form of log line data and syntactical attributes crucial for anomaly detection. Unsupervised clustering algorithms (K-Means, DBScan, and EM clustering) evaluate this approach using three datasets: ISOT-CID, ECML/PKDD 2007, and an Indonesian Apache Access datasets. The syntactic approach outperforms traditional log abstraction techniques, consistently disambiguating malicious behaviour in various traffic log data forms. The highest performance is achieved with K-Means clustering, with accuracy scores ranging from 0.6-0.7 for anomaly detection across the three datasets.

Expanding on the previously discussed work, Karlsen et al. \cite{CSNet} continue the exploration of syntactic feature extraction with application log files and compare it against a semantic approach. The semantic approach utilizes an LLM-based feature extraction method which extracts a sentence embedding representation from the raw logs. They utilize the DistilRoBERTa model and perform a comparison of this new method against the short syntactic methodology outlined by Copstein et al. \cite{copstein} using four unsupervised machine learning algorithms on three datasets, ECML/PKDD 2007, CSIC 2010, and the Indonesian Apache Access datasets. The resulting study notes that the semantic approach outperforms that of the short syntactic method by about 5\% and has the most potential for further development.

\subsection{LLM-Based Approaches}

In recent work, Nam et al. \cite{vmBERT} explore BERT for log analysis in Virtual Machine (VM) failure prediction. They propose a model where BERT extracts sentence embeddings for each log, which are then fed into a pre-trained convolutional neural network for predicting failures within a 2 to 30-minute window. They achieve an accuracy of 0.74 on an in-house OpenStack VM system log test and training set, which is not publicly available. A notable benefit of this BERT-CNN model is that the embedding representation enables the prediction of VM failure even in cases not observed during training.

Wang et al. \cite{Wang} propose an intrusion detection system that utilizes word2vec and TF-IDF as a feature extraction method with supervised methods\footnote{Gradiant boosted decision tree, Naive Bayes classifier, and LSTM} and is benchmarked for classification. This system is evaluated on the supercomputer system log datasets with the best results indicating that TF-IDF/word2vec extraction methods and LSTM perform with an accuracy of 0.99. 

Research by Qi et al. \cite{Qi} proposes another deep learning-based anomaly detection system called Adanomaly which performs anomaly detection using a combination of log key extraction using the Drain log parser (\cite{Drain}), and an adversarial autoencoder method for reconstructive error-based anomaly detection. They evaluate this model using three system log files with an accuracy changing between 0.92 - 0.98. However, the use of the log parser, Drain, to extract log key representations reduces the generalizeability of the approach.

Recently, Seyyar et al. \cite{Seyyar} proposed an intrusion detection system specifically for web applications that first extracts the URL from the application logs, and then generates an embedding using an NLP transformer model - BERT. This provides the embedding for a CNN classifier for performing intrusion detection. Inference speeds of 0.4ms are reported at an average accuracy of 0.96 when evaluated on the CSIC, FWAF, and HTTPparams datasets. However, this system performs URL feature extraction as part of the pre-processing step, a process that limits its generalization.

Guo et al. \cite{guo2021logbert} propose an anomaly detection system called LogBERT. LogBERT leverages the masked language modelling feature of the NLP transformer model BERT to perform a self-supervised form of anomaly detection on pre-parsed logs extracted using Drain. The model is designed to identify key features in the log representation, thus making predictions about what feature will appear next. The LogBERT system is applied to system logs such as BGL and Thunderbird system datasets and the model reports accuracies of 0.8232, 0.9083, and 0.9664 respectively. Notably, this method also relies on the rule-based log parsing methodology of Drain to extract meaningful log key representations, i.e. limited to scenarios for which appropriate rules exist.

Shao et al. \cite{Shao} propose an improvement on the LogBERT model called Prog-BERT-LSTM. This builds on LogBERT by adapting the sequence feature learning components with an LSTM. The semantic understanding of log sequence is improved, and as a result, when applied to the BGL and Thunderbird systems datasets they are able to report accuracies from approximately 0.84 to 0.97. Notably, this method also relies on the Drain rule-based log parsing methodology to extract meaningful log key representations and hence inherits the same limitations.

A study by Guo et al. \cite{guotranslog} proposes a novel supervised transformer-based pipeline for log anomaly detection called TransLog. The proposed system leverages the enhanced sentence embeddings of the sentence-BERT LLM to vectorize pre-processed log data in the form of log templates extracted using Drain. The embeddings of the log templates are used to pre-train a transformer encoder model in which a classification head is employed for the task of anomaly detection. They add a `light adapter' to the transformer architecture to enable faster fine-tuning through targeted training of specific components of the pre-trained model. This approach is evaluated using publicly available system logs such BGL, and Thunderbird achieving F1-scores around0.99. Naturally, this approach is reliant on pre-processed system logs due to the use of Drain to extract log template representations and so relies on a static log parser.

Le et al. \cite{Le} attempt to incorporate the BERT language model as an approach to vectorizing pre-processed log lines. This representation is then fed into a transformer encoder block which alongside a softmax layer classifies anomalous events. The resulting model named NeuralLog is evaluated on the BGL, Thunderbird, HDFS, and Spirit datasets. This system aims to replace the out-of-vocabulary limitations of typical log parsing approaches that struggle to adapt to new words in the log file, i.e. Out Of Vocabulary words (OOV). The resulting model provides performance around an F1-score of 0.96. It should be noted that this approach still requires pre-processing of the log files in order to encode and classify.  

Sec2vec is an NLP transformer-based method for web log vectorization proposed by Gniewkowski et al. \cite{Sec2Vec} in 2023. This system uses the RoBERTa language model to perform log feature extraction, through fine-tuning of both the original model and its tokenizer. Additionally, the tokenizer has had its tokenization strategy swapped for a Byte-level Byte Pair Encoder in order to benefit from improved token-level representations with Byte-level encoding representations of the underlying text. As part of the process for preparation, this system is fine-tuned using the log file. Afterwards, an embedding representation can be extracted from the same log file through the concatenation of the hidden states from the final four layers of the LLM. This embedding representation of the log is evaluated using the Random Forest classifier. Benchmarking is performed using web service, and URL datasets specifying the vector size as 3072. Sec2Vec returns F1-scores (in average) around 0.98. It should be noted that an embedding space of 3072 is verbose and while the study does not report computational costs for any of the steps, fine-tuning, training, and performing inference with this vector space will be expensive. Additionally, they do not perform an ablation study, so it is not clear how much the computationally intensive steps contribute to the performance.

\subsection{Summary}

The literature presents several gaps in NLP-based log analysis, as follows. 
\begin{itemize}
    \item Log analysis research typically explores either web application logs with the publicly avaialable CSIC, and PKDD datasets, or they evaluate performance on publicly avaialable system logs such as Thunderbird, and BGL (Table \ref{tbl:lit_llm_results}). In practice, systems will be deployed under both settings, so should be benchmarked on both types of data.
    \item The LLMs employed focus on a limited set of LLM architectures, Table \ref{tbl:llm_type}. However, the literature for state-of-the-art transformer language models provides several alternatives to the BERT/RoBERTa models that have unique differences or improvements that have yet to be explored in log analysis.
    \item Performance is only assessed in terms of the modified LLM. As such it is difficult to determine how much performance improvement results from the `optimization' versus the original LLM.
\end{itemize}

Thus, in this paper, our goal is to explore the versatility of LLM-based models for log analysis with several language model architectures and evaluate them using the original language models as well as the fine-tuned language models to understand the effects (if any) and requirements of the fine-tuning process. The results of this study provide a benchmark and foundation for LLM-based feature extraction in log analysis for intrusion detection. 

\begin{table}
\caption{F1-score Results from Literature}\label{tbl:lit_llm_results}
\centering
\begin{tabular}{lccccc}
\toprule
Study       & CSIC   & PKDD   & Thunderbird & BGL    & Spirit    \\
\midrule
Nguyen \cite{supe}*   & 0.9098 & 0.9256 & -           & -                 & -      \\ 
Bhatnager \cite{supe2}*  & 0.996  & 0.996  & -           & -                 & -      \\ 
Wang \cite{Wang}       & -      & -      & 0.99        & -                 & -      \\ 
Seyyar \cite{Seyyar}    & 0.96   & -      & -           & -                 & -      \\ 
Gniewkowski \cite{Sec2Vec} & 0.98   & -      & -           & -                 & -      \\ 
Farzad  \cite{farzad}*    & -      & -      & 0.993       & 0.994             & -      \\ 
Hongchen Guo \cite{guotranslog} & -      & -      & 0.998       & 0.98              & -      \\ 
Le \cite{Le}         & -      & -      & 0.96        & 0.98              & 0.97   \\ 
Sivri \cite{sivri}      & 0.982  & -      & -           & -                 & -      \\ 
Adhikari \cite{adhikari}*   & 0.93   & -      & -           & -                 & -      \\ 
\botrule
\end{tabular}
\smallskip
\textit{*Indicates a result in which cited work reports Accuracy rather than F1-score.}
\end{table}

\begin{table}
\caption{Overview of Supervised Methodologies Used in Previous LLM Studies, RegEx indicates that useful features were handpicked and extracted from the logs}\label{tbl:llm_type}
\centering
\begin{tabular}{lccc}
\toprule
Study       & Pre-parsed     & Traditional NLP & LLM-Based \\
\midrule
Nguyen \cite{supe}      & -              & YES             & NO              \\
Bhatnager  \cite{supe2} & -              & YES             & NO              \\
Farzad \cite{farzad} & -              & YES              & NO              \\
Wang   \cite{Wang}     & -              & YES             & NO              \\
Sivri \cite{sivri} & RegEx             & YES              & NO              \\
Adhikari \cite{adhikari} & RegEx              & YES              & NO              \\
Nam   \cite{Seyyar}   & RegEx          & NO              & BERT            \\
Seyyar   \cite{Seyyar}   & URL extraction & NO              & BERT            \\
Gniewkowski \cite{Sec2Vec} & -              & NO              & RoBERTa         \\ 
Hongchen Guo \cite{guotranslog} & Drain              & NO              & BERT           \\
Le \cite{Le} & Drain              & NO              & BERT            \\
\botrule
\end{tabular}
\end{table}

\section{Methodology}\label{Approach}
In the following, we first summarize the various properties of the datasets adopted for benchmarking (Section \ref{sec:datasets}). Section \ref{sec:workflow} introduces the LLM4Sec pipeline\footnote{The code for the LLM4Sec pipeline will be made publicly available upon acceptance.} as a workflow applied to log files encompassing a portfolio of LLM (Section \ref{sec:llm_summary}), visualization routines (Section \ref{sec:visuals}) and performance metrics. Such a workflow is generic in that no additional pre- or post-processing steps are deployed.

\subsection{Datasets}\label{sec:datasets}

This study explores the efficacy of a general approach to log analysis which leverages the flexibility of feature engineering approaches in semantic analysis on six log files, three are web application logs, and three are system logs. Since a primary component of this research is to explore a general approach to application and system log analysis for security purposes, it is important to employ data from different sources. Therefore,
six publicly available datasets are employed in this research. Each dataset represents a different real-life system as summarized below, in Table \ref{table1} and \ref{table2}. For the purposes of this research, the labels were reduced to a binary characterization of the task. Thus, classes associated with malicious behaviour were all labelled as ANOMALOUS and safe activity was labelled as NORMAL. The details of each dataset are described as follows:

\begin{table}[htbp]
\caption{Application Logs}
\centering
\begin{tabular}{lcccccc}
\label{table1}
      & \multicolumn{2}{c}{AA} & \multicolumn{2}{c}{CSIC} & \multicolumn{2}{c}{PKDD} \\
Class & \# & \% & \# & \% & \# & \% \\ 
\midrule
Normal       & 29,778 & 85.71\% & 36,000 & 59.01\% & 35,006 & 69.85\% \\
Anomalous    & 4,965 & 14.29\% & 25,000 & 30.99\% & 15,109 & 30.15\% \\ 
\midrule
Total        & 34,743 & 100\% & 61,000 & 100\% & 50,115 & 100\% \\ 
\botrule
\end{tabular}
\end{table}

\begin{table}[htbp]
\centering
\caption{NLP Characteristics of the application log datasets}\label{nlpapplication}
\begin{tabular}{lcccccc}
\toprule
Dataset            & Word Length & Word Count & Character Count \\
\midrule
AA                 & 12.22       & 16.85      & 222.13          \\
Standard Deviation & 2.87        & 1.83       & 46.07           \\
\midrule
CSIC               & 15.94       & 27.55      & 467.02          \\
Standard Deviation & 3.50        & 0.90       & 104.13          \\
\midrule
PKDD               & 11.49       & 92.1       & 1124.25         \\
Standard Deviation & 0.21        & 0.50       & 3.28            \\
\botrule
\end{tabular}
\end{table}

\begin{table}[h]
\centering
\caption{System Log properties}

\begin{tabular}{lcccccc}
\label{table2}
      & \multicolumn{2}{c}{Thunderbird} & \multicolumn{2}{c}{Spirit} & \multicolumn{2}{c}{BGL} \\
Class & \# & \% & \# & \% & \# & \% \\
\midrule
Normal       & 207,963,953 & 98.46\% & 90,656,272 & 33.29\% & 4,399,503 & 92.66\% \\
Anomalous    & 3,248,239 & 1.54\% & 181,642,697 & 66.71\% & 348,460 & 7.34\% \\ 
\midrule
Total        & 211,212,192 & 100\% & 272,298,969 & 100\% & 4,747,963 & 100\% \\  
\botrule
\end{tabular}
\end{table}

\begin{table}[htbp]
\centering
\caption{NLP Characteristics of the system log datasets}\label{nlpsystem}
\begin{tabular}{lcccccc}
\toprule
Dataset            & Word Length & Word Count & Character Count \\
\midrule
Thunderbird               & 8.51       & 15.48       & 146.62         \\
Standard Deviation & 3.19        & 5.32       & 69.41            \\
\midrule
Spirit                 & 7.00       & 17.40      & 137.53          \\
Standard Deviation & 0.89        & 2.73       & 22.26           \\
\midrule
BGL               & 10.18       & 15.34      & 160.30          \\
Standard Deviation & 1.62        & 8.36       & 56.48          \\
\botrule
\end{tabular}
\end{table}

\subsubsection{Application Logs}

Application logs, combining service-specific log files and often network traffic, provide a comprehensive view of application and network-level interactions for security-related events. The Apache Access Dataset, published in 2022 by Indramayu State Polytechnic University, focuses on intrusion detection in web server access logs, including 37,693 records of normal, suspicious, and attack queries, representing typical non-anonymized application logs. Hereafter, this dataset is referred to as AA \cite{AA}. The CSIC Dataset, introduced by the Information Security Institute of the Spanish Research National Council in 2010, serves as an updated resource for intrusion detection research in web application logs. The dataset simulates an e-commerce environment with various attack types like SQL injection and buffer overflow and is notable for its non-anonymized data \cite{CSIC}. Finally, the ECML/PKDD Dataset from the 2007 Discovery Challenge, referred to as PKDD, comprises 50,115 log records with HTTP packet details for anomaly detection, including a wide range of attack behaviours like XSS and SQLi, characterized by its anonymized data for privacy, making it the most verbose dataset in terms of log line representation \cite{PKDD}.

\subsubsection{System Logs}

A set of datasets was published in 2007 by Oliner et al. \cite{Tbird} that consists of a series of labelled system/event logs produced by supercomputer clusters in June 2006. Events are labelled with alert and non-alert tags. Each dataset is named after the supercomputer from which they originated and the three datasets used in this work are Thunderbird, BGL, and Spirit. These logs differ significantly from the previous application-style logs in that they are a record of processor-level activity, and are significantly more volumous as seen in Table \ref{table2}.

\subsection{LLM4Sec Pipeline}\label{sec:workflow}

\begin{figure}
    \centering
    \includegraphics[width=1\linewidth]{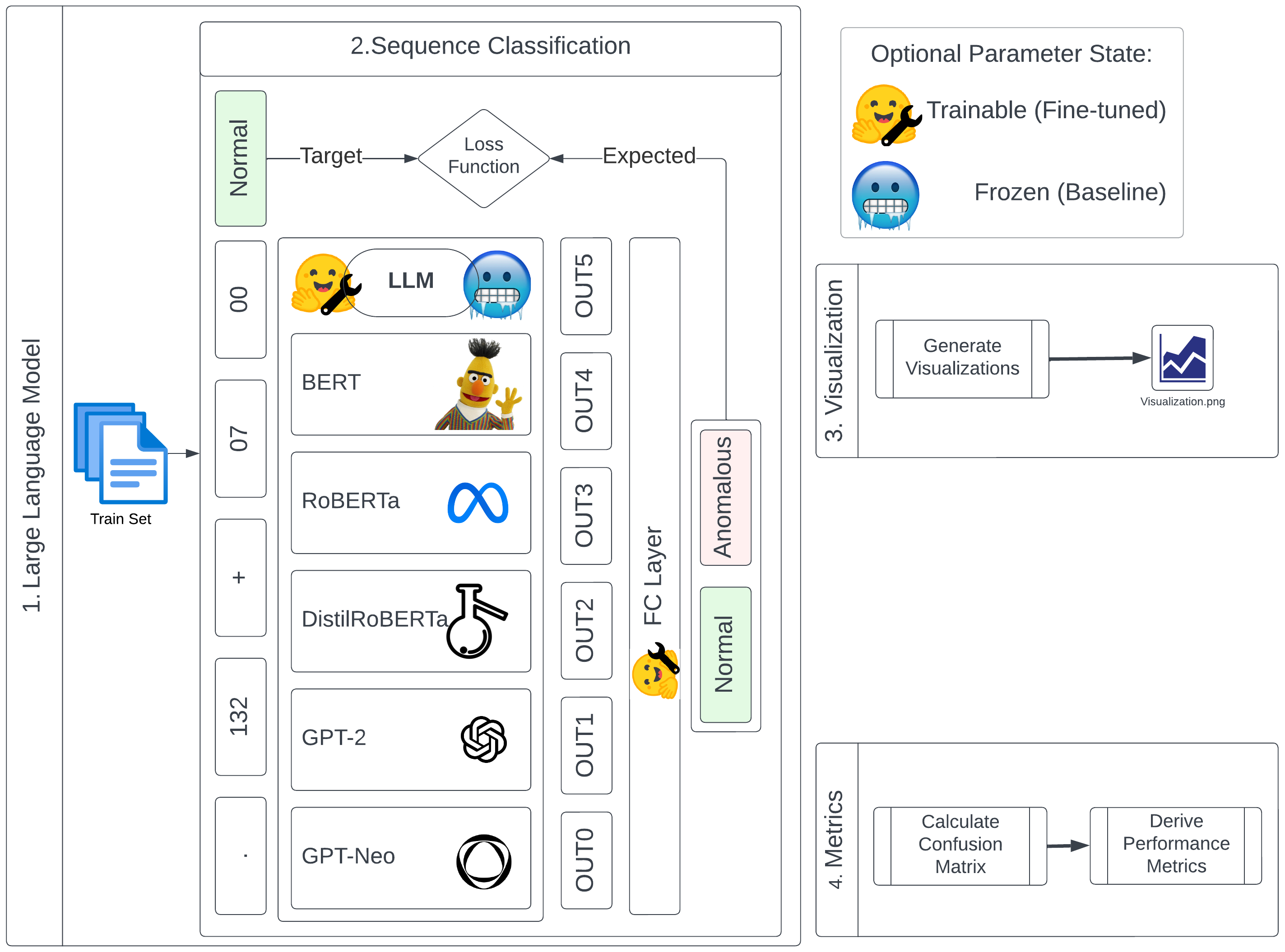}
    \caption{LLM4Sec pipeline overview. The original LLM architecture is augmented with a fully connected (FC) layer for the purpose of anomaly detection}
    \label{SecLLM}
\end{figure}

As part of this research, a new benchmarking experimentation pipeline is developed in order to streamline the research conducted in this study. LLM4Sec is a flexible and componentized CLI tool for log analysis that enables researchers to easily evaluate the performance of different LLMs for the task of log analysis. Integration of methods for interpretation and visualization, and fine-tuning options provide the necessary utilities to expand and continue this research. LLM4Sec deploys the same sequence of `modules' that the experimental structure of this study follows: fine-tuning, visualizations, and performance analysis, Figure \ref{SecLLM}. The LLM-related elements assume the Huggingface platform\footnote{\url{https://huggingface.co}}, supporting a wide range of models. In addition to the LLM-based feature extraction methods, syntactic approaches such as TF-IDF, and the statistical NLP approach provided by Copstein et al. are also available for experimentation \cite{copstein}.

\subsection{Large Language Models}\label{sec:llm_summary}

This module of the LLM4Sec pipeline, shown in Figure \ref{SecLLM}, utilizes the log data to fine-tune the pre-trained LLMs and generate vectorized log representations that embed the semantic relations between the tokens in the log. As the first module in the pipeline, it accepts the raw log file as input, then divides the log file into 70\% training, 10\% validation and 20\% testing, and provides log-specific LLMs and a vectorized log representation as the output feeding into the subsequent module. The details of the LLMs evaluated in this research are provided in the following subsections. 

\subsubsection{BERT}

\begin{figure}[h]%
\centering 
\includegraphics[width=0.9\textwidth]{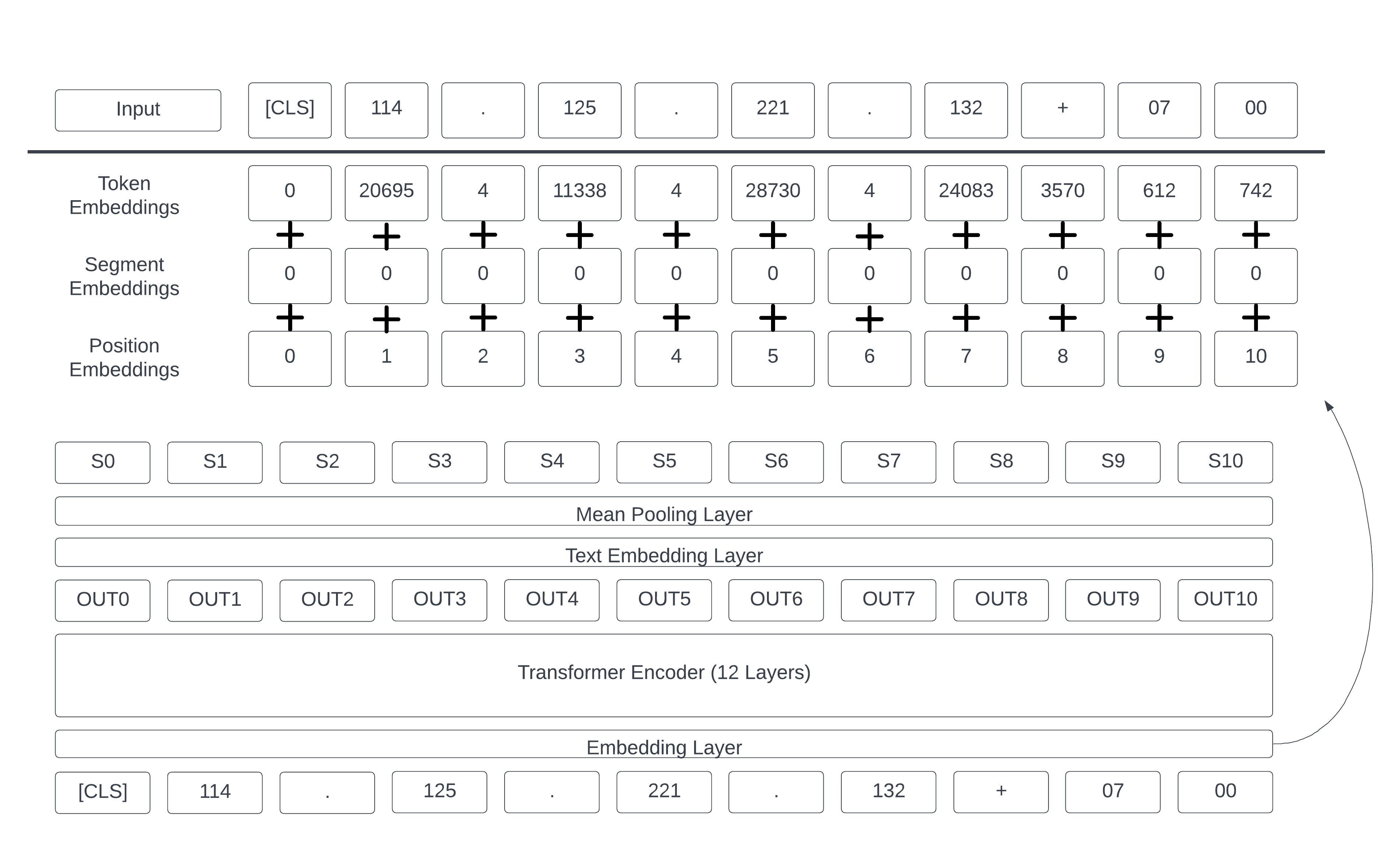}
\caption{This figure shows the sentence embedding extraction technique applied to a CSIC log line using BERT}\label{BERT}
\end{figure}

The BERT model (Bidirectional Encoder Representations from Transformers) revolutionized NLP upon its release by Google in 2018, employing a bidirectional approach to the interpretation of textual data based on the transformer architecture introduced by Vaswani et al. \cite{transformer}. Trained on an expansive corpus of 16GB, including over 3.3 billion words from Wikipedia and the Brown Corpus, BERT's architecture of 12 transformer encoder layers, each with 768 hidden states, leverages multi-head self-attention and feed-forward layers for contextual learning of words in a sentence. Tokenization is managed by the WordPiece algorithm \cite{Wordpiece}, which breaks down words into recognizable subunits for the model. BERT's training utilizes masked language modelling—masking 15\% of words for context-based prediction—and next-sentence prediction, enhancing its sequential understanding. These techniques enable fine-tuning for specific domains, as shown in the LLM4Sec fine-tuning module in Figure \ref{SecLLM}. In order to generate meaningful embedding representations of logs we employ sentence embedding vectorization. This is accomplished by performing mean pooling on the final layer of the LLM to reduce the logits to a 768-dimension vector. The process is outlined in the model's architectural visualization for sentence embedding extraction in Figure \ref{BERT}.

\subsubsection{Roberta}

The RoBERTa model, developed by Facebook AI in 2019, improves on the original BERT architecture with several key additions, earning its name as the Robustly Optimized BERT Approach \cite{RoBERTa}. It was trained on a massive 160GB text corpus, extending beyond BERT's 16GB dataset with additional web and news content. Notably, RoBERTa omits the Next-sentence Prediction (NSP) task, a decision informed by evidence that NSP was not crucial to BERT's success. RoBERTa also adopts a different pre-training configuration, utilizing larger batch sizes and fewer steps to achieve lower perplexity and quicker training. Unlike BERT's static approach, RoBERTa introduces dynamic masking during training, where the masked tokens vary during the learning process. Lastly, it uses Byte Pair Encoding (BPE) for tokenization, which results in a more extensive vocabulary of over 50,000 tokens compared to BERT's 30,000, facilitating a more nuanced understanding of language. This LLM provided improved performance over the BERT model in many areas of NLP, motivating its inclusion in this evaluation of LLM performance in log analysis.

\subsubsection{DistilRoberta}

In 2019, Sanh et al. introduced the concept of distillation in NLP, a method to streamline large BERT-style models for efficiency, as seen in their study on distilRoBERTa \cite{DistilRoBERTa}. This technique uses a teacher-student model where a smaller ``student'' model learns to replicate the larger ``teacher'' model's outputs. The student's training focuses on minimizing the difference between its outputs and the teacher's, effectively transferring knowledge. Applied to models like BERT and RoBERTa, distillation reduces the layers, parameters, and training time significantly, resulting in faster inference times. DistilRoBERTa, for instance, uses 6 layers and 82 million parameters, compared to RoBERTa's 12 layers and 125 million parameters, halving the inference time while maintaining output fidelity. This model was selected as the smallest of the LLMs being evaluated, it aims to provide the robust performance of RoBERTa in half the computation cost, which is particularly important in the case of log analysis where inference time is a significant factor in the setting of real-time intrusion detection.

\subsubsection{GPT-2}

OpenAI's GPT-2, released in 2019, is a departure from BERT-style models, with its unique autoregressive, decoder-only transformer architecture designed for translation, text generation, and question answering \cite{GPT2}. It is uni-directional, masking future tokens during training to prevent the model from using future context, unlike the bi-directional approach of BERT. The model uses self-attention, encoder-decoder attention, and feed-forward layers for contextual processing and next-token prediction. Trained on a diverse 40GB dataset from the web, GPT-2's full version has 1.5 billion parameters, though a smaller variant with 124 million parameters was used for this research \cite{GPT2}. It shares the BPE tokenization method with RoBERTa, having a similar-sized vocabulary of over 50,000 tokens. To facilitate sentence embedding extraction like BERT, padding tokens are introduced to GPT-2, aligning it with BERT's method for fixed-length sentence embeddings by mapping a padding token to the existing eos\_token in the tokenizer. This LLM was selected to provide a contrasting perspective to that of the encoder-only transformer LLMs, providing log analysis results from a completely different architecture.

\begin{figure}[h]%
\centering
\includegraphics[width=0.9\textwidth]{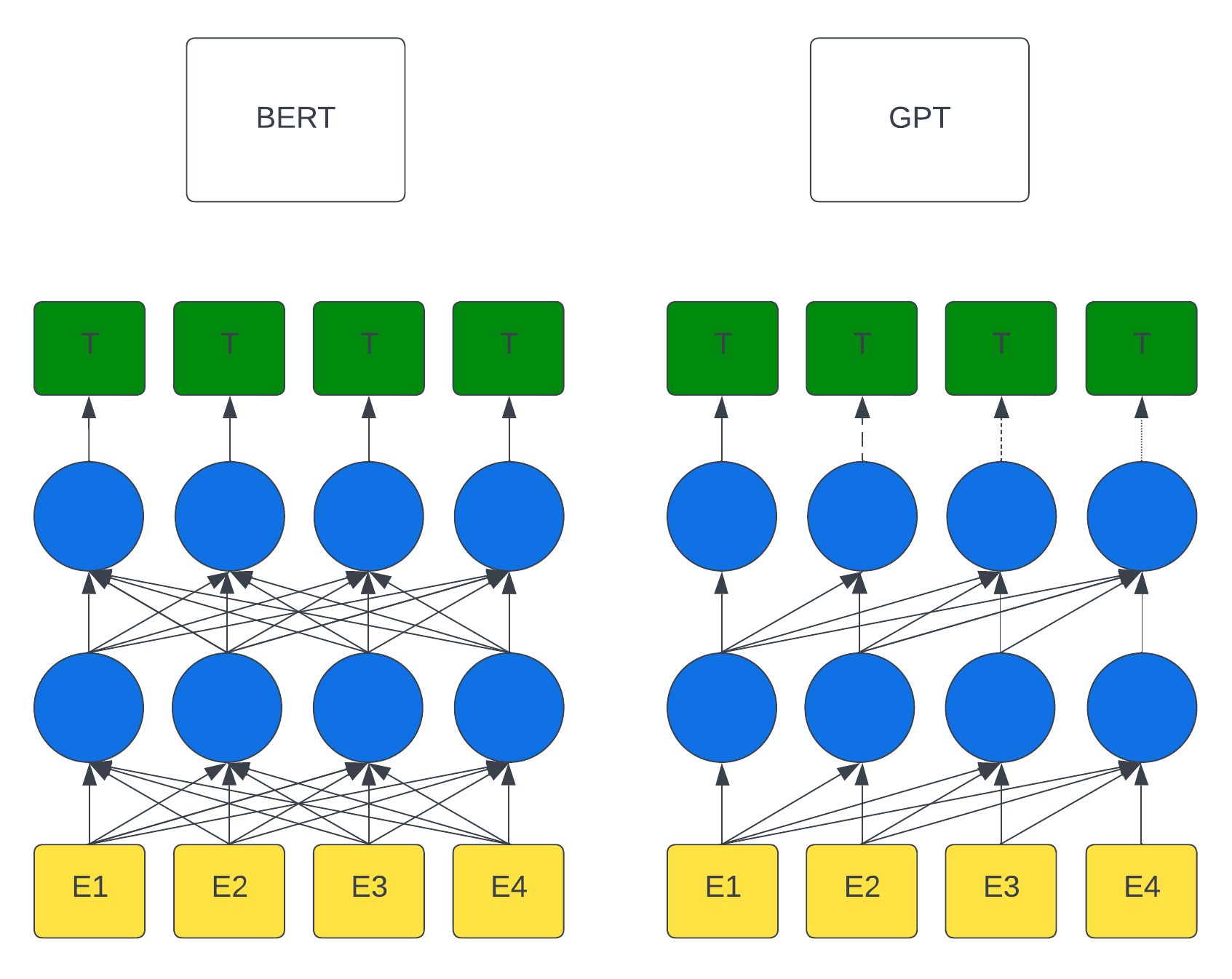}
\caption{A simplified visualization of how the encoder-based architecture of BERT differs from the decoder-based architecture of GPT}\label{BERTvsGPT}
\end{figure}

\subsubsection{GPT-Neo}

The GPT-Neo model is an Eleuther AI replication of the closed source GPT-3 model produced by Open AI and as such is an autoregressive decoder style language model \cite{GPTNEO}. This model also assumes a casual language model and utilizes a slightly modified autoregressive decoder-style architecture to that of GPT-2. GPT-Neo differs from GPT-2 in that it utilizes local attention between every other layer, and was trained on an 825 GB dataset or `the pile'. Training using the pile exposes the LLM to a wide variety of different text sources including books, GitHub repositories, webpages, chat logs, as well as research papers from a number of STEM fields. The GPT-Neo model comes in a variety of different model sizes as well ranging from 2.7 billion parameters to 125 million. For the purposes of this research, the 1.3 billion parameter model was utilized. Evaluating this model's performance in log analysis will not only further expand the range of different LLMs being evaluated, but will also enable an efficiency analysis of the model with more than 1 billion parameters.\footnote{All other models have just over 100 million parameters.}

\begin{table}
		\begin{tabular}{lcccc}
                \toprule
			Language Model & Word Embedding Space & Parameters & Training Data Size \\
			\midrule
			Bert-base-cased & 768 & 110M & 16 GB \\
			\midrule
			RoBERTa & 768 & 123M & 160 GB \\
			\midrule
			DistilRoBERTa-base & 768 & 82M & 40 GB \\
			\midrule
			GPT-2 & 768 & 124M & 40 GB \\
			\midrule
			GPT-Neo & 2048 & 1.3B & 800 GB \\
            \botrule
		\end{tabular}
	\caption{Model Characteristic of the Different Large Language Models employed}
	\label{LLM}
\end{table}

\subsection{LLM Sequence Classification}

The sequence classification module seen in Figure \ref{SecLLM} shows the adaptation of an LLM for sequence classification. For sequence classification tasks using models like BERT, a dense fully connected linear layer is typically added after the final layer of the model. This then uses the output of the final transformer layer to map to a probability distribution for classification categories. This process, which employs supervised learning, requires labelled data to fine-tune the linear layer's and the language model's weights for precise classification, and adopts a loss function such as cross-entropy:

\begin{equation}\label{crossentropy}
L = -\frac{1}{N} \sum_{i=1}^{N} \sum_{c=1}^{C} y_{i,c} \log(p_{i,c})
\end{equation}
\noindent Where:
\begin{itemize}
    \item \( N \) is the total number of samples in the dataset.
    \item \( C \) is the number of classes in the classification problem.
    \item \( y_{i,c} \) is a binary indicator (0 or 1) if class label \( c \) is the correct classification for observation \( i \).
    \item \( p_{i,c} \) is the predicted probability that observation \( i \) belongs to class \( c \).
\end{itemize}

Cross-entropy quantifies the difference between the predicted probability distribution \( p_{i,c} \) and the actual distribution \( y_{i,c} \), guiding the model's training by minimizing this discrepancy. During fine-tuning, the model is initialized with a pre-trained tokenizer, equipped with a linear classification layer, and configured with specific training parameters (optimizer, learning rate, and loss function). Fine-tuning uses the model's inherent training objective—masked language modelling for BERT-like models or causal language modelling for GPT-like models. Although comprehensive fine-tuning can be computationally demanding, an efficient approach is to freeze the pre-trained layers, refining only the classification layer to reduce computational demands significantly.

\subsection{Interpretation and Visualization}\label{sec:visuals}

The Interpretation and Visualization module of the LLM4Sec pipeline is responsible for generating visualizations that improve the explainability of the LLMs. This module accepts the vector log embedding representation as input, as well as the sequence classifier's predicted labels, and true labels in order to feed the t-SNE, and SHAP algorithms utilized in this module. The t-SNE plot is generated from the full dataset while the SHAP visualization enables log-specific explainability. An overview of this module is seen in Figure \ref{SecLLM}.

\subsubsection{SHAP}

TranSHAP, developed by Kokalj et al., \cite{kokalj} extends the Kernel SHAP method to explain the decision-making process of transformer-based language models. Kernel SHAP itself is an interpretable machine learning approach, which approximates Shapley values to determine feature importance for any given model prediction. TranSHAP adapts this to transformer models by first fragmenting input sequences into indexable parts, and then creating perturbed versions of these sequences to analyze. It uses the model's tokenizer and classifier to predict outcomes of perturbed inputs and feeds these back into the Kernel SHAP algorithm, which calculates the importance of each subword fragment. Despite the computational intensity, TranSHAP can provide valuable insights into how language models weigh individual input components during classification.

\subsubsection{t-SNE}

The t-Distributed Stochastic Neighborhood Embedding (t-SNE) is a non-linear technique used for dimensionality reduction. It is particularly effective in visualizing high-dimensional datasets albeit at high computational cost \cite{tSNEOG}. Pairwise similarities are first calculated in the high-dimensional space, after which the data is mapped onto a low-dimensional space aiming to preserve these similarities. This is achieved through iterative minimization of the Kullback-Leibler divergence between the two spaces using gradient descent, optimizing the map to reveal the data's intrinsic structure in a way that is particularly informative for visual analysis.

\section{Evaluations and Results}\label{sec2}

\subsection{Metrics}

The performance metrics adopted in this study include F1-score, Precision, and Recall. Since all the datasets have labels provided by the authors, these metrics were calculated post-training and post-testing to measure the performance of each evaluation. The metrics were derived from the confusion matrices of the training and test evaluations for each technique used in the study. 

The F1-score is a traditional method for measuring the model performance on a particular dataset, it is computed using Precision and Recall. Precision gives an indication of the percentage of true positives (in class) which are relevant while Recall gives an indication of the percentage of true positives out of all the true positives available in the data. The formulae for Precision, Recall and F1-score are given in the following equations \ref{P}, \ref{R} and \ref{F1} \cite{metrics}.

\begin{equation}\label{P}
P= \frac{TP}{TP+FP}
\end{equation}
        
\begin{equation}\label{R}
R= \frac{TP}{TP+FN}
\end{equation}

\begin{equation}\label{F1}
F1 = \frac{2PR}{P+R}
\end{equation}

Weighted F1-Score is used to calculate the per class weighted mean of all F1-scores, that is to say, the average F1-score where per class proportion, S, of the dataset is considered. This is calculated with the following formulae in Eq. \ref{weighted}.

\begin{equation}\label{weighted}
F1_{weighted}= \sum\frac{F1_{class}}{S_{class}}
\end{equation}

False positive rate (FPR) is used to calculate the rate at which a model classifies negatively labelled data as positive, particularly important in the case of intrusion detection because these would imply that ANOMALOUS security events were classified as NORMAL. This is calculated in the following formulae in Eq. \eqref{fpr_formula}

\begin{equation}\label{fpr_formula}
FPR = \frac{FP}{FP + TN}
\end{equation}

Further metrics that are of importance to consider for these models are the computational cost of inference and the cost of performing fine-tuning. Specifically, application and systems logs are fast-growing with high velocity, so inference has to keep up with new events, and fine-tuning LLMs can be a costly endeavour, possibly delaying model deployment.

\subsection{Hyperparameters}

The two components of the LLM4Sec where hyperparameter tuning occurs are in the LLM module. In this section, hyperparameters are specified to improve the quality of classification and learning of log representations. The hyperparameters for the fine-tuning of LLMs with the sequence classification objective are outlined in Table \ref{hyperparams2}. These represent the recommended language modelling settings for fine-tuning LLMs.

\begin{table}
\caption{Hyperparameters employed in this research}
\centering
\begin{tabular}{lc}
\toprule
\textbf{Hyperparameter} & \textbf{Value} \\
\midrule
Learning Rate & $2 \times 10^{-5}$ \\
Train Batch Size & 8 \\
Eval Batch Size & 8 \\
Seed & 42 \\
Optimizer & Adam with betas=$(0.9, 0.999)$ and epsilon=$1 \times 10^{-8}$ \\
LR Scheduler Type & Linear \\
Number of Epochs & 10 (BERT, RoBERTa, DistilRoBERTa, GPT-2) 3 (GPT-Neo)\\
\bottomrule
\end{tabular}
\label{hyperparams2}
\end{table}

\subsection{Efficiency Analysis: Feature Extraction and Fine-tuning}

In this research, we fine-tune each of the five chosen Language Model Models (LLMs) using sequence classification tasks on a common computing pipeline (Table \ref{table:simplified_machine_specs}). This fine-tuning process is computationally intensive, and the training times are detailed in Table \ref{FTSup}. Notably, the largest model, GPT-Neo, incurs significantly higher fine-tuning costs compared to the others. This is because GPT-Neo is over 10 times the size of the other models.

Across all experiments, DistilRoBERTa consistently exhibits the shortest fine-tuning times. This is because the DistilRoBERTa model is the smallest in model size relative to the other models with only 82 million parameters.

With regards to the dataset-specific fine-tuning times, we note that PKDD takes significantly longer than the other datasets due to its verbose log line length see Table \ref{nlpapplication}. Alternately, Spirit provides the fastest of the fine-tuning results and this is due to its comparatively smaller average log line length see Table \ref{nlpsystem}.

A clear efficiency trend is seen in sequence classification inference times shown in Table \ref{INFSup}, where GPT-Neo shows high inference times, and this highlights the advantages of the smaller DistilRoBERTa in terms of computational efficiency.

\begin{table}
\caption{Benchmarking Pipeline Specification}
\label{table:simplified_machine_specs}
\centering
\begin{tabular}{lc}
\toprule
\textbf{Component} & \textbf{Specification} \\
\midrule
CPU & Intel Xeon Silver 4214 CPU @ 2.20GHz \\
CPU Architecture & Intel x86 \\
Number of Cores & 6 cores \\
RAM & 80 GB \\
Storage & 160 GB \\
GPU & NVIDIA A100 \\
GPU Architecture & NVIDIA Ampere Architecture \\
\bottomrule
\end{tabular}
\end{table}

\begin{table}
    \caption{Fine-tuning time for sequence classification (minutes)}\label{FTSup}
		\begin{tabular}{llcccccc}
                \toprule
			LLM & Type & AA & CSIC & PKDD & Thunderbird & Spirit & BGL \\
			\midrule
			\multirow{2}{*}{DistilRoBERTa} & Baseline & 13.287 & 27.550 & 32.245 & 20.557 & 9.409 & 17.261 \\
			& Fine-tuned & 28.871 & 73.859 & 74.059 & 46.886 & 20.158 & 39.794 \\
			\midrule
                \multirow{2}{*}{BERT} & Baseline & 24.220 & 58.663 & 58.763 & 45.976 & 17.893 & 36.275 \\
			& Fine-tuned & 56.688 & 73.859 & 141.858 & 46.886 & 42.380 & 88.245 \\
			\midrule
                \multirow{2}{*}{RoBERTa} & Baseline & 22.855 & 57.331 & 57.387 & 36.075 & 15.651 & 30.614 \\
			& Fine-tuned & 53.868 & 140.459 & 140.659 & 88.378 & 36.663 & 75.058 \\
			\midrule
                \multirow{2}{*}{GPT-2} & Baseline & 30.780 & 44.056 & 135.931 & 18.981 & 11.178 & 18.370 \\
			& Fine-tuned & 41.936 & 134.599 & 320.879 & 63.658 & 29.881 & 51.659 \\
			\midrule
                \multirow{2}{*}{GPT-Neo} & Baseline & 172.161 & 449.350 & 833.166 & 172.228 & 138.661 & 173.293 \\
			& Fine-tuned & 229.815 & 366.570 & 862.830 & 177.735 & 134.145 & 181.980 \\
                \botrule
		\end{tabular}
\end{table}

\begin{table}
    \caption{Inference time for Sequence Classification LLM experiments (logs per second)}
    \label{INFSup}
    \begin{tabular}{lcccccc}
        \toprule
        LLM & AA & CSIC & PKDD & Thunderbird & Spirit & BGL \\
        \midrule
        BERT & 76.400 & 77.800 & 71.300 & 76.300 & 76.000 & 77.000 \\
        DistilRoBERTa & 133.900 & 135.800 & 119.600 & 135.100 & 135.700 & 135.100 \\
        GPT-2 & 65.500 & 68.800 & 63.600 & 65.900 & 65.900 & 66.000 \\
        GPT-Neo & 27.300 & 19.600 & 10.600 & 32.700 & 33.200 & 30.200 \\
        RoBERTa & 76.700 & 80.100 & 73.100 & 76.500 & 76.500 & 76.300 \\
        \botrule
    \end{tabular}
\end{table}

\begin{table}
    \caption{Testing F1-Scores for Sequence Classification LLM with and without fine-tuning. Higher values are better}
    \label{SupF1}
    \begin{tabular}{llccccccr}
        \toprule
        LLM & Type & AA & CSIC & PKDD & TB & Spirit & BGL & Average \\
        \midrule
        \multirow{3}{*}{DistilRoBERTa} & Baseline & 0.929 & 0.696 & 0.844 & 1.000 & 0.992 & 0.985 & \textbf{0.908} \\
        & Fine-tuned & 1.000 & 0.996 & \textbf{0.991} & 1.000 & 1 & 1.000 & \textbf{0.998} \\
        & Change & \textbf{0.071} & \textbf{0.300} & \textbf{0.147} & 0 & \textbf{0.008} & \textbf{0.015} & \textbf{0.090} \\
        \midrule
        \multirow{3}{*}{RoBERTa} & Baseline & 0.832 & 0.640 & 0.677 & 0.999 & 0.987 & 0.955 & 0.848 \\
        & Fine-tuned & 0.999 & \textbf{0.998} & 0.984 & 1.000 & 1 & 1 & 0.997 \\
        & Change & \textbf{0.167} & \textbf{0.357} & \textbf{0.307} & \textbf{0.001} & \textbf{0.013} & \textbf{0.045} & \textbf{0.148} \\
        \midrule
        \multirow{3}{*}{BERT} & Baseline & 0.786 & 0.530 & 0.575 & 0.977 & 0.875 & 0.898 & 0.773 \\
        & Fine-tuned & 1 & 0.998 & 0.966 & 1.000 & 1 & 1 & 0.994 \\
        & Change & \textbf{0.214} & \textbf{0.468} & \textbf{0.391} & \textbf{0.023} & \textbf{0.125} & \textbf{0.102} & \textbf{0.220} \\
        \midrule
        \multirow{3}{*}{GPT-Neo} & Baseline & 0.926 & 0.700 & 0.673 & 1.000 & 0.991 & 0.996 & 0.881 \\
        & Fine-tuned & 0.992 & 0.994 & 0.985 & 1.000 & 0.995 & 0.958 & 0.987 \\
        & Change & \textbf{0.066} & \textbf{0.294} & \textbf{0.312} & 0.000 & \textbf{0.005} & -0.038 & \textbf{0.107} \\
        \midrule
        \multirow{3}{*}{GPT-2} & Baseline & 0.827 & 0.238 & 0.581 & 0.975 & 0.534 & 0.898 & 0.676 \\
        & Fine-tuned & 0.833 & 0.992 & 0.959 & 0.983 & 0.993 & 0.956 & 0.953 \\
        & Change & \textbf{0.006} & \textbf{0.754} & \textbf{0.378} & \textbf{0.008} & \textbf{0.459} & \textbf{0.057} & \textbf{0.277} \\
        \botrule
    \end{tabular}
\end{table}

\begin{table}
    \caption{False positive rates with and without fine-tuning. Lower values are better}
    \label{FPRSup}
    \begin{tabular}{llcccccc}
        \toprule
        LLM & Type & AA & CSIC & PKDD & Thunderbird & Spirit & BGL \\
        \midrule
        \multirow{2}{*}{DistilRoBERTa} & Baseline & 0.407 & 0.538 & 0.391 & \textbf{0.006} & 0.009 & 0.209 \\
        & Fine-tuned & \textbf{0} & 0.008 & \textbf{0.023} & \textbf{0.006} & \textbf{0} & \textbf{0.002} \\
        \midrule
        \multirow{2}{*}{RoBERTa} & Baseline & 0.857 & 0.640 & 0.835 & 0.084 & 0.007 & 0.553 \\
        & Fine-tuned & \textbf{0} & 0.004 & 0.045 & \textbf{0.006} & \textbf{0} & \textbf{0} \\
        \midrule
        \multirow{2}{*}{BERT} & Baseline & 1 & 0.861 & 1.000 & 1 & 0.000 & 1 \\
        & Fine-tuned & \textbf{0} & 0.004 & 0.096 & 0.006 & \textbf{0} & \textbf{0} \\
        \midrule
        \multirow{2}{*}{GPT-Neo} & Baseline & 0.335 & 0.192 & 0.696 & \textbf{0.006} & 0.005 & 0.041 \\
        & Fine-tuned & 0.013 & 0.014 & 0.035 & \textbf{0.006} & 0.004 & 0.502 \\
        \midrule
        \multirow{2}{*}{GPT-2} & Baseline & 0.875 & \textbf{0.002} & 0.756 & 1 & \textbf{0} & 1 \\
        & Fine-tuned & 0.005 & 0.016 & 0.121 & 0.006 & 0.005 & 0.530 \\
        \botrule
    \end{tabular}
\end{table}

\subsection{Results}

The results, presented in Table \ref{SupF1}, illuminate the impact of performing fine-tuning on LLMs for sequence classification with a particular application for security tasks. The data reveal testing F1-Scores for supervised sequence classification models that have been employed across the five LLMs (Section \ref{sec:llm_summary}), tested against six distinct datasets (Section \ref{sec:datasets}).

The results highlight that fine-tuning is not only significant but also consistently observed across models. This improvement is particularly pertinent to the precision required in intrusion detection, where the F1-Scores of classification are integral to the model's loss function during fine-tuning. On the other hand, baseline performances, where the training is confined to the classification head alone, often fall short, sometimes even completely failing to classify any anomalous event. It is also apparent that the DistilRoBERTa model provides a better baseline and post-fine-tuning F1-Scores for sequence classification tasks within the security domain.

When inspecting the false positive rates in Table \ref{FPRSup}, it becomes apparent that without comprehensive fine-tuning of the underlying LLM, the ability to classify sequences accurately for intrusion detection is significantly compromised, frequently mistaking ANOMALOUS activities for NORMAL ones. It is noteworthy that models such as DistilRoBERTa, RoBERTa, and BERT not only excel but consistently deliver some of the most accurate classification results seen in the literature for intrusion detection, achieving near or absolute-perfect scores.

These findings emphasize the critical importance of fine-tuning in supervised learning for security and reaffirm the advantageous performance of BERT-style models in such high-stakes security applications.

\subsection{Visualization}

In this section, we explore the visualizations available under different LLMs. In the ideal case, the LLM-based feature sets should be both discriminatory and interpretable. 

\begin{figure}[h]%
\centering
\includegraphics[width=0.9\textwidth]{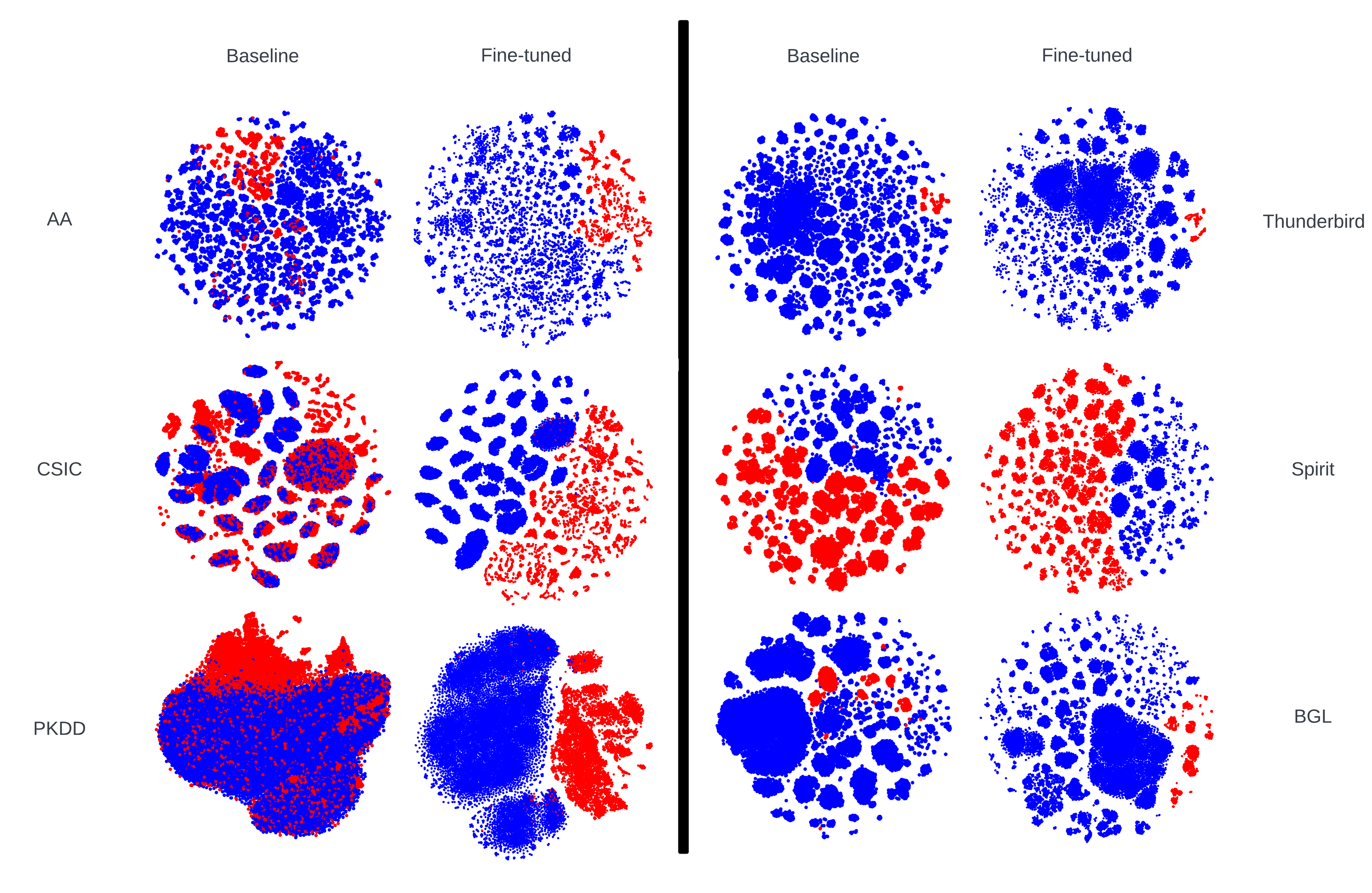}
\caption{t-SNE visualizations of the embeddings produced by baseline and fine-tuned DistilRoBERTa models for all six datasets. Left (Right) pairs represent Application (System) logs}\label{tSNEDroBERTa}
\end{figure}

The t-SNE plots were generated for the log embedding representations produced by the baseline and fine-tuned DistilRoBERTa models and can be seen in figure \ref{tSNEDroBERTa}. Inspection of these plots shows effective separation of NORMAL and ANOMALOUS logs particularly in the system logs. While the difference between the baseline and fine-tuned models is not immediately obvious, some characteristics are shared across the visualizations which indicate some improvements in the model's ability to effectively understand the logs. Fine-tuned t-SNE plots appear to show less mixed behaviour in the clusters, and more concrete lines of separation between NORMAL, and ANOMALOUS groups. Even the PKDD dataset which is characterized by high degrees of randomness due to its anonymization sees some improvement in the disambiguation of abnormal events.

\begin{figure}[h!]%
\centering
\includegraphics[width=0.9\textwidth]{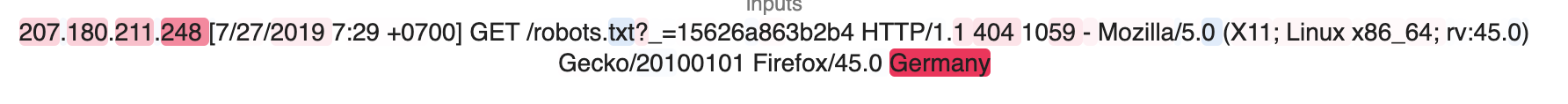}
\caption{SHAP explanation for log line from AA dataset using Finetuned DistilRoBERTa model}\label{AA_SHAP}
\end{figure}

\begin{figure}[h!]%
\centering
\includegraphics[width=0.9\textwidth]{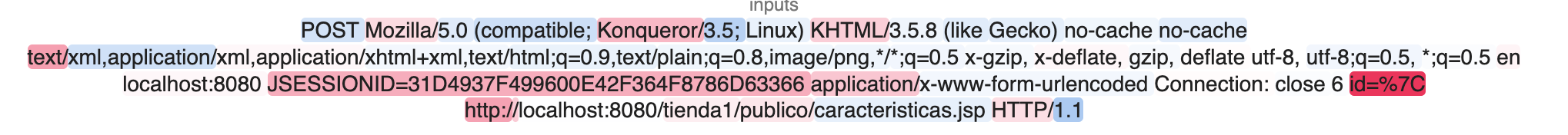}
\caption{SHAP explanation for first example log line from CSIC dataset using Finetuned DistilRoBERTa model}\label{CSIC_SHAP1}
\end{figure}

\begin{figure}[h!]%
\centering
\includegraphics[width=0.9\textwidth]{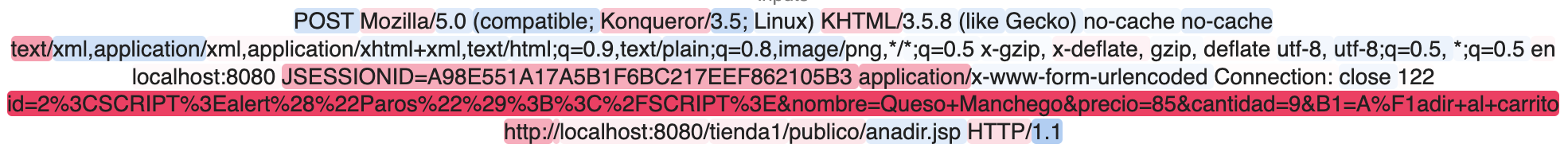}
\caption{SHAP explanation for second example of log line from CSIC dataset using Finetuned DistilRoBERTa model}\label{CSIC_SHAP2}
\end{figure}

\begin{figure}[h!]%
\centering
\includegraphics[width=0.9\textwidth]{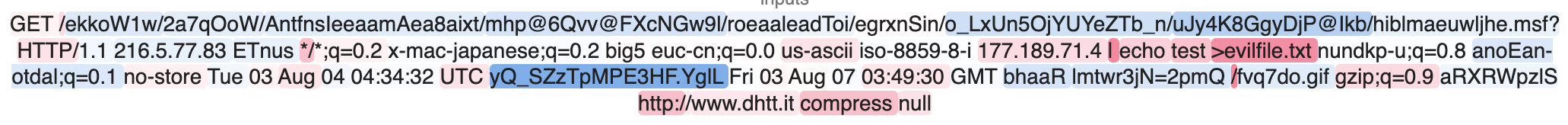}
\caption{SHAP explanation for first example of log line from PKDD dataset using Finetuned DistilRoBERTa model}\label{PKDD_SHAP1}
\end{figure}

\begin{figure}[h!]%
\centering
\includegraphics[width=0.9\textwidth]{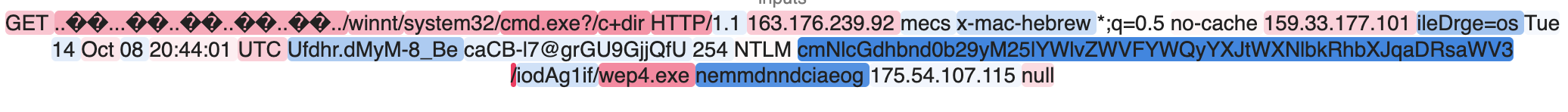}
\caption{SHAP explanation for second example of log line from PKDD dataset using Finetuned DistilRoBERTa model}\label{PKDD_SHAP2}
\end{figure}

\begin{figure}[h!]%
\centering
\includegraphics[width=0.9\textwidth]{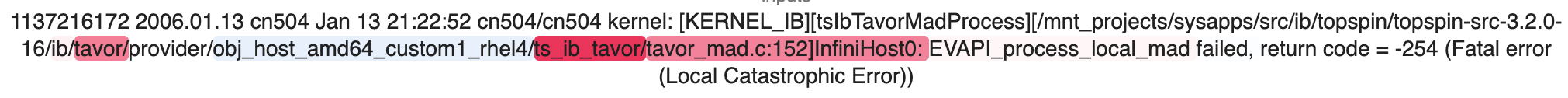}
\caption{SHAP explanation for log line from Thunderbird dataset using Finetuned DistilRoBERTa model}\label{TB_SHAP}
\end{figure}

\begin{figure}[h!]%
\centering
\includegraphics[width=0.9\textwidth]{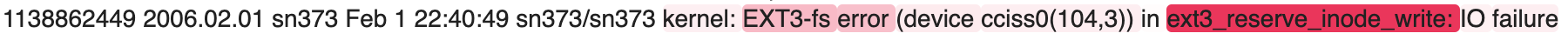}
\caption{SHAP explanation for log line from Spirit dataset using Finetuned DistilRoBERTa model}\label{Spirit_SHAP}
\end{figure}

\begin{figure}[h!]%
\centering
\includegraphics[width=0.9\textwidth]{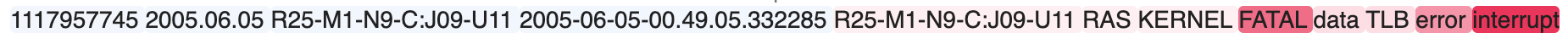}
\caption{SHAP explanation for log line from BGL dataset using Finetuned DistilRoBERTa model}\label{BGL_SHAP}
\end{figure}

The SHAP plots were generated using unique samples from each dataset on the DistilRoBERTa model from the supervised experiments. Inspection of the different SHAP plots produced by the fine-tuned DistilRoBERTa models shows the training for sequence classification has promoted automatic feature learning in the classification models. The attention mechanisms in the LLM have been tuned to identify unique patterns present in the NORMAL and ANOMALOUS logs. Properties of notable significance include:

\begin{itemize}
    \item AA: As seen in figure \ref{AA_SHAP} the language model has identified the significance of individual IP addresses, location, and malformed URLs. These are features that a security expert may also evaluate in the identification of security incidents, particularly in the case of Apache web server logs.
    \item CSIC: Figures \ref{CSIC_SHAP1} and \ref{CSIC_SHAP2} illustrate how more advanced attack types are being identified by the unique characteristics of the log. The model has identified specific log characteristics associated with the use of path traversing or RCE, exploits. Even more notably one log line is flagged as anomalous for having malicious ID values submitted to the service. 
    \item PKDD: Figures \ref{PKDD_SHAP1} and \ref{PKDD_SHAP2} illustrate that even though this dataset is anonymized exploits characterized by invalid file access, SQL injection, and path traversal are found in the URL and parameters have been identified by the model. Even misused credentials are being identified by the supplied parameters in this model. 
    \item Thunderbird: Figure \ref{TB_SHAP} shows that the model has adapted to identify log characteristics associated with kernel panic or error codes. These are learned features associated with abnormal events.
    \item Spirit: Figure \ref{Spirit_SHAP} shows that the model has identified precise kernel error codes associated with failures similar to that of Thunderbird.
    \item  BGL: Figure \ref{BGL_SHAP} illustrates results, in this case, the model associates the keyword 'FATAL', and 'INTERRUPT' with anomalous behaviour.
\end{itemize}

In-depth inspection of the visualizations provided by both the embeddings from the visualizations from t-SNE, and SHAP shows that the DistilRoBERTa model has captured the structure and meaning of each of the logs after fine-tuning. The pattern identification shown in the outputs of SHAP analysis indicates that the models' attention mechanisms have identified important aspects of the log line. It also shows that appropriate attention has been given to clear markers of ANOMALOUS security events like XSS, LDAP Injection, etc. This automatic feature learning results from fine-tuning the model for sequence classification. Subsequent analysis through SHAP would be of value to a security analyst since important parts of the log file which deserve further exploration are highlighted for the security analyst. 

\subsection{Discussion}

In examining the performance of various LLM approaches for log analysis, especially when applying domain adaptation for sequence classification, distinct advantages emerge. Notably, the DistilRoBERTa model, upon fine-tuning, achieves near-perfect classification performance across all six datasets. This performance significantly exceeds both its baseline variant and other LLMs. Such results underscore the significance of fine-tuning, as exemplified in the CSIC dataset, where all tested LLMs experienced performance improvements ranging from 0.3 to 0.75 in terms of F1-Scores. This evidence firmly establishes the critical role of fine-tuning in optimizing LLMs for detailed and accurate log analysis tasks.

Additionally, the tranSHAP visualizations, which provide an in-depth analysis of the specific elements in a log line that identify it as an anomaly, could be highly beneficial in security analysis, particularly in cyber forensics. When examining the visualizations generated by the DistilRoBERTa model (Figures \ref{AA_SHAP} to \ref{BGL_SHAP}), the method of highlighting log characteristics is able to explicitly identify the nature of the anomalies. An excellent illustration of this is seen in Figure \ref{CSIC_SHAP1}, where the intensity of red highlighting effectively pinpoints a Cross-Site Scripting (XSS) attack as the most significant component of the log. This feature demonstrates the potential of an LLM-based approach to log analysis as a powerful tool for interpretable and accurate identification of security threats.

This study has uncovered notable benefits of using LLM-based methods for anomaly detection. One of the main advantages is the ability to produce highly interpretable results through the use of tranSHAP, coupled with a high classification performance. DistilRoBERTa, when fine-tuned for sequence classification, emerges as the most efficient model, while also offering competitive classification performance that places it among the top performing supervised methods in the field, Table \ref{SOTA}. This approach demonstrates equal or better results compared to the existing literature in four out of five previously studied datasets, with the AA dataset being a new area of exploration. However, this high performance comes at a computational cost. Despite DistilRoBERTa nearly halving the time required for fine-tuning and inference, it remains resource-intensive, achieving a maximum classification rate of 135 logs per second and requiring over an hour for fine-tuning. While efficiency is often overlooked in the literature and challenging to benchmark due to varied experimental setups, it remains a critical factor. This is not only for ensuring swift response times to security threats in enterprise environments but also for considering the energy requirement implications of this method.

\begin{table}
\caption{Comparison with earlier benchmarking studies in F1-Score. Previous results are limited to either a subset of Application or System Log files}\label{SOTA}
\centering
\begin{tabular}{lccccc}
\toprule
Study       & CSIC   & PKDD   & Thunderbird & BGL    & Spirit    \\
\midrule
Nguyen \cite{supe}*   & 0.9098 & 0.9256 & -           & -                 & -      \\ 
Bhatnager \cite{supe2}*  & 0.996  & 0.996  & -           & -                 & -      \\ 
Wang \cite{Wang}       & -      & -      & 0.99        & -                 & -      \\ 
Seyyar \cite{Seyyar}    & 0.96   & -      & -           & -                 & -      \\ 
Gniewkowski \cite{Sec2Vec} & 0.98   & -      & -           & -                 & -      \\ 
Farzad  \cite{farzad}*    & -      & -      & 0.993       & 0.994             & -      \\ 
Hongchen Guo \cite{guotranslog} & -      & -      & 0.998       & 0.98              & -      \\ 
Le \cite{Le}         & -      & -      & 0.96        & 0.98              & 0.97   \\ 
Sivri \cite{sivri}      & 0.982  & -      & -           & -                 & -      \\ 
Adhikari \cite{adhikari}*   & 0.93   & -      & -           & -                 & -      \\ 
\midrule
LLM4Sec    & \textbf{0.998}   & 0.991      & \textbf{1.0}      & \textbf{1.0}               & \textbf{1.0}      \\
\botrule
\end{tabular}
\smallskip
\textit{*Indicates a result in which cited work reports Accuracy rather than F1-score.}
\end{table}

\section{Conclusion}

This research has provided a benchmark for five LLMs over six datasets for the task of log analysis in security using sequence classification. Specifically, 60 fine-tuned language models for log analysis are deployed and bechmarked.The research provides compelling results for the continued exploration of LLM-based log analysis given that the results of the sequence classification approach outperform the state-of-the-art log analysis across all datasets, especially in the case of DistilRoBERTa. 

The benefits of fine-tuning can also be seen clearly in the results where it consistently provides positive improvement across all six datasets. In the case of sequence classification, the improvements range from 1 to 20\% improvement. It is clear that domain adaptation of LLMs that were originally trained on text corpus can provide important improvements in the language model's ability to correctly interpret the significance of components of the log.

In addition, the opportunities for clearly legible and interpretable visualization through the t-SNE and SHAP analysis cannot be understated. The opportunity to highlight and present the components of the log line which contributes most to their classification as NORMAL or ANOMALOUS using SHAP is unique to this LLM-based approach. Such a visualization would enable security experts to see clearly why a log is considered outside of normal. Furthermore, both t-SNE and SHAP analysis visualizations show the advantages of performing fine-tuning for such analysis.

Finally, the extensive experimentation discussed throughout this research was accomplished through the use of the proposed LLM and NLP based pipeline, LLM4Sec. It is our aim that LLM4Sec enables research with LLM and NLP techniques in log analysis, for not only the reproducibility of the research presented in this paper, but also expanding on it.

Future work will continue exploring the impact of adopting different tokenization methods for the LLMs, as well as pre-training, which historically has led to  enhanced domain-specific performance. Additionally, training a new security-specific LLM from scratch with an appropriate corpus of log data using the foundational work accomplished in this study might provide much-needed efficiency improvements. Specifically, the model could afford to be much smaller with less emphasis on typical textual language corpa. Such a security-specific LLM would require an appropriate corpus of diverse log files. This would in and of itself be a novel task since there are no publicly available security logs that could provide insight into all layers of a system's architecture.

\bibliography{bibplz}

\end{document}